# Directly accessing octave-spanning dissipative Kerr soliton frequency combs in an AlN microring resonator

Haizhong Weng[1,2,†], Jia Liu[3,†], Adnan Ali Afridi[1,2], Jing Li[1,2], Jiangnan Dai[3], Xiang Ma[3], Yi Zhang[3], Qiaoyin Lu[3],

John F. Donegan[1,2]*, and Weihua Guo[3]*

[1]CRANN and AMBER research centers, Trinity College Dublin, Dublin, Dublin 2, Ireland

[2]School of Physics, Trinity College Dublin, Dublin, Dublin 2, Ireland

[3]Wuhan National Laboratory for Optoelectronics, and School of Optical and Electronic Information, Huazhong University of Science and Technology, 1037 Luoyu Road, Wuhan 430074, China

[†]These authors contributed equally: Haizhong Weng, Jia Liu.

*Correspondence and requests for materials should be addressed to J.F.D (email: jdonegan@tcd.ie) or W.H.G (email: guow@mail.hust.edu.cn)

## Abstract

Self-referenced dissipative Kerr solitons (DKSs) based on optical microresonators offer prominent characteristics including miniaturization, low power consumption, broad spectral range and inherent coherence for various applications such as precision measurement, communications, microwave photonics, and astronomical spectrometer calibration. To date, octave-spanning DKSs with a free spectral range (FSR) of ~1 THz have been achieved only in ultrahigh-$Q$ silicon nitride microresonators, with elaborate wavelength control required. Here we demonstrate an octave-spanning DKS in an aluminium nitride (AlN) microresonator with moderate loaded $Q$ ($5\times10^5$) and FSR of 374 GHz. In the design, a $TE_{00}$ mode and a $TE_{10}$ mode are nearly degenerate and act as pump and auxiliary modes. The presence of the auxiliary resonance balances the thermal dragging effect in dissipative soliton comb formation, crucially simplifying the DKS generation with a single pump and leading to a wide single soliton access window. We experimentally demonstrate stable DKS operation with a record single soliton step (~80 pm) and octave-spanning bandwidth (1100-2300 nm) through adiabatic pump tuning and on-chip power of 340 mW. Our scheme also allows for direct creation of the DKS state with high probability and without elaborate wavelength or power schemes being required to stabilize the soliton behavior.

**Introduction**

Over the past few decades, optical frequency combs based on mode-locked lasers have been proved to be a revolutionary technology[1] that enables various applications in optical clock[2], frequency metrology[3], astrophysical spectrometer calibration[4], and arbitrary waveform generation[5]. Kerr frequency comb (microcomb) generation based on parametric four-wave mixing (FWM) in monolithic microresonators with high quality ($Q$) factors[6,7], emerged as an alternative scheme, has attracted considerable interest due to miniaturization, chip-scale integration, repetition rate in the microwave and terahertz regime, and spectral coverage from visible to midinfrared[8-11]. The recent demonstration of dissipative Kerr solitons (DKSs)[12], with a double balance between Kerr nonlinearity and dispersion as well as loss and parametric gain in the optical resonator, have provided a route to a fully coherent microcomb with a smooth spectral envelope and broadened width due to soliton-induced Cherenkov radiation[13,14]. The soliton microcomb has ubiquitous commercial potential for applications in dual-comb spectroscopy[15], coherent communications[16], frequency synthesizer[17], ultra-fast distance measurements[18], and microwave photonics[19].

However, accessing soliton states in microresonators is still challenging since the DKS generation requires keeping the pump in an effective red-detuned regime where the thermo-optic instability in the microresonator causes complex behaviour[12]. Solitons have been demonstrated in silica[10,20], crystalline $MgF_2$[12,21], silicon nitride ($Si_3N_4$)[13,21-23], aluminum nitride (AlN)[24], and lithium niobite ($LiNbO_3$)[25,26] based systems. Sundry techniques were adopted to demonstrate the above-mentioned solitons including different frequency tuning schemes such as rapid forward or/and backward pump laser scans, fast control of the resonator temperature, as well as pump power modulation and power kicking. However, any complicated control or extra equipment needed will increase the cost and limit the practical applications of solitons. Moreover, single soliton operation in the octave-spanning regime, a pre-requisite for the $f$–$2f$ self-referencing technique for determining the carrier-envelope offset frequency, has only been demonstrated in a $Si_3N_4$ microresonator with ultra-high $Q$ factors and large free spectral range (FSR) of ~1 THz[17,27,28].

An alternative method to manipulate the power coupled into the resonance is using an auxiliary laser to pump another cavity mode, which can suppress thermal dragging dynamics in dissipative soliton comb formation and increase the soliton stability and access window[29-32]. However, generally for $Si_3N_4$ and AlN microresonators, both pump and auxiliary lasers need to be amplified to high powers with two commercial fiber amplifiers. Similarly, when an auxiliary mode has another polarization and is located at a slight red-detuning of the pump mode[27], a single soliton state with a narrow step width of ~2 pm has been accessed in $Si_3N_4$ with a single pump. AlN has a similar optical refractive index and Kerr nonlinear coefficient as $Si_3N_4$, as well as a wide transparency range, and has been used for microcomb generation[33,34]. In addition, the strong Pockels ($\chi^2$) nonlinear effect in AlN has been exploited for frequency doubling, tripling and bi-chromatic microcomb generation[35,36]. Recently, Pockels quadratic soliton microcombs ranging from 1380 to 1750 nm, with low comb threshold and high pump-to-soliton conversion efficiency, have been realized by pumping an AlN microresonator at 780 nm[37]. Nevertheless, accessing the Kerr soliton state with a single pump and a direct laser tuning method hasn't been reported for this material.

In this work, we reveal that an adjacent mode near the pump resonance can help to mitigate thermo-optical effects, thereby allowing the access to octave-spanning DKS generation in an AlN microresonator (FSR ~374 GHz) with a single pump. The microresonator was designed meticulously to realize two transverse-electrical (TE) mode families near 1550 nm and then we pumped the fundamental TE ($TE_{00}$) mode, while the adjacent first-order TE ($TE_{10}$) mode on the red-detuned side (~35 pm mode separation) can mitigate the thermal requirements for accessing soliton states. By choosing appropriate power, a stable single soliton state (with a broad step of ~80 pm, 10 GHz) with an octave-spanning spectral bandwidth can be accessed at a slow pump tuning speed of 1 nm/s. Owing to the wide soliton step, we could directly access the single DKS states, with high probability, by simply switching the pump wavelength from an off- to an on-resonance state with one step. The soliton pulse width was measured to be ~30 fs using an autocorrelation measurement, indicating high coherence of the soliton microcomb. The scheme was further demonstrated in another device with a different mode separation experimentally (~12 pm, see Supplementary Fig. S4). The proposed technique is an important step forward and paves the way to enabling robust and stable self-referenced single soliton operation in AlN microresonators as well as other materials.

## Results

**Device design, characterization and technique**

Figure 1 illustrates the device characterization and the approach taken in this work. Using standard photolithography and inductively coupled plasma etching processes (see in methods), we fabricated devices in a quarter of a 2 inch wafer (Fig. 1b), that consists of a 1.2-μm-thick epitaxial single-crystal AlN film and a sapphire substrate[38,39]. In this work, we employed a microring resonator with a radius of 60 μm, a waveguide width of 2.29 μm (Fig. 1c), and the height is targeted at 1.2 μm (fully etched). The integrated bus waveguide, 500 nm apart from the resonator, is tapered from 0.91 to ~4 μm at both ends to reduce the fiber-waveguide coupling loss, which is measured to be 3.2 dB per facet. Figure 1d shows the experimental transmission trace for a fabricated device. Two sets of TE polarization modes, $TE_{00}$ (FSR ~374 GHz) and $TE_{10}$ (FSR ~366 GHz) are easily identified due to the different FSRs and marked by blue and red lines, respectively. The two close modes around 1550.7 nm have a separation of only 35 pm, as the zoomed-in view shown in Fig. 1e, while the $TE_{00}$ mode on the blue-detuned side is used for pumping subsequently. The loaded $Q$ factor ($Q_L$) and intrinsic $Q$ factor ($Q_{int}$) for the pump mode are $4.8\times10^5$ and $1.6\times10^6$, respectively, corresponding to a propagation loss of 0.23 dB/cm. The $Q_L$ and $Q_{int}$ for the nearby $TE_{10}$ mode are $2.8\times10^5$ and $7.5\times10^5$, respectively. The threshold powers for parametric oscillation for the $TE_{00}$ and $TE_{10}$ modes are calculated to be 39 and 145 mW (see methods). For the same resonator geometry, we have observed the similar behavior (two close TE modes in the communication C band) in multiple devices, indicating that the design is easy to replicate in fabrication.

Besides the Kerr nonlinearity, another basic requirement for achieving broadband Kerr solitons is near-zero anomalous dispersion. In the context of microresonator-based Kerr frequency combs, the dispersion properties can be extracted from the Taylor-expanded resonance frequencies with a central pump frequency $\omega_0$:

$$\omega_\mu = \omega_0 + \mu D_1 + \frac{1}{2!}\mu^2 D_2 + \frac{1}{3!}\mu^3 D_3 + \frac{1}{4!}\mu^4 D_4 + \cdots \qquad (1)$$

Where $\mu$ is the mode number deviation from the center pump, $\omega_\mu$ is the angular frequency at mode number $\mu$, $D_1/2\pi$ is the FSR around $\omega_0$, and $D_2/2\pi$ is second-order dispersion, relative to the group velocity dispersion (GVD)[12]. For DKS generation, it is generally required to achieve anomalous GVD in the microresonator ($D_2 > 0$). Since AlN bulk material has normal dispersion, here we compensated the material dispersion with the geometry-dependent waveguide dispersion by designing the ring resonator structure. The target geometry was chosen based on finite element method (FEM) simulations, in order to ensure anomalous GVD for the $TE_{00}$ mode and that its resonance frequency is slightly higher than that of $TE_{10}$ mode (several GHz, see Supplementary Fig. S2). The integrated dispersion defined by $D_{int} = \omega_\mu - (\omega_0 + \mu D_1) = \sum_{i\geq 2}\frac{1}{i!}\mu^i D_i, i \in N$ was calculated and plotted in Fig. 1f. We set the center pump frequency $\omega_0/2\pi$ to 193.292 and 193.289 THz for $TE_{00}$ and $TE_{10}$ modes, respectively. Their $D_1/2\pi$ are 374.2 and 368.6 GHz, while the second-order dispersion $D_2/2\pi$ are extracted to be 4.8 and 36.8 MHz, indicating both mode families are in anomalous dispersion, while the $TE_{00}$ mode has near-zero dispersion and the $D_{int} = 0$ is obtained around 255.4 THz (~1170 nm). We successfully observed the DKS by pumping the $TE_{00}$ mode, featuring a greater $Q$ factor and flatter engineered dispersion than those of $TE_{10}$ mode, assisted by the adjacent $TE_{10}$ mode (see the principle in Fig. 1g). The auxiliary $TE_{10}$ mode can compensate for the intracavity power change when the soliton is generated and mitigate the thermal requirements to access stable soliton states (see Fig. 1h).

**Octave-spanning dissipative Kerr solitons generation**

In experiment, we can reach the solitons states by sweeping the pump wavelength under suitable powers, allowing the balancing of thermal and optical effects are needed for DKS formation. Figure 2b shows the pump transmission after the fiber Bragg-grating (FBG), at 350 mW on-chip power, when sweeping the laser wavelength over the resonance at 1nm/s speed. Clearly, a 67-pm-range step-like structure characteristic of soliton formation is formed before entering the $TE_{10}$ resonance. By sweeping the laser from 1550.55 nm to different detuning positions and recording the frequency comb spectra, we observe that the stable comb spectrum changes from primary comb state (see Supplementary Fig. S3) to modulation instability (MI, Fig. 2c, white dash line i) comb state and soliton states (Fig. 2c, white dash lines ii and iii at different detuning points), which are plotted in Fig. 2d. The solitons exist stably when the wavelength stops between 1550.776 and 1550.831 nm (i.e., soliton step length 55 pm), consistent with the transmission in Fig. 2b. Upon further tuning the pump toward the red-sided regime before dropping out of the cavity resonance, we can only observe a few weak sidebands generated from the $TE_{10}$ mode (see Supplementary Fig. S3) due to the lower $Q$ factor and larger dispersion. Therefore, our geometry design is ideal for generating the DKSs without any concern about the nonlinear competition between the two close modes. Figure 2d-i shows the generated MI comb spectrum, which ranges from 140 to 260 THz and has a dispersive wave (DW) like bump around 255 THz. Typical frequency combs of soliton 1 and soliton 2 can be found in Figs. 2d-ii and -iii, which cover an octave-spanning range from 130 to 273 THz (1100-2300 nm). The soliton centroids are 5.2 and 7.4 THz lower than the pump, which is caused by the Raman self-frequency shift effect[40,41]. Similar to the predicted spectra obtained from numerical simulation (see Fig. 5), the measured solitons spectra are significantly extended towards shorter wavelengths due to the emission of the DW via soliton-induced Cherenkov radiation at ~264 THz (1136 nm). By fitting the spectral envelopes with $sech^2$ shape (dash lines), we can find the full-width half-maximums (FWHMs) of soliton 1 and soliton 2 are 15.2 and 12.3 THz, corresponding to 40 and 33 modes, respectively. The minimum soliton pulse width $\tau$ is estimated to be 21 and 26 fs for soliton 1 and 2, using 0.315×the pulse FWHM where the pulse shape is $sech^2$ [12]. The transition from MI comb state to the soliton regime has been verified by the drastic reduction of low-frequency intensity noise (see Fig. 2e). There is a sharp signal at ~ 3 GHz for soliton 2, but still implies a low-noise state of the frequency comb. To further confirm the single DKS states, the temporal characteristics of the solitons were also carried out through the second-harmonic generation (SHG) autocorrelation (AC) measurement. The pump was suppressed by ~26 dB through an FBG filter, and a fiber polarization controller (FPC) was applied before the autocorrelator. Figure. 2f shows the measured AC traces separated by the cavity round-trip time of 2.66 ps, inversely proportional to the FSR of ~374 GHz. The trace of soliton state is much noisier than that of the MI comb state as the soliton comb power is at the limit that can be detected by the autocorrelator. Within a narrow AC measurement range, an extremely narrow pulse with ~30 fs width (Fig. 2g, soliton 2 has similar pulse shape and width) can be obtained for soliton 1 based on the $sech^2$ fitting, which is close to the estimated minimum transform-limited pulse width 21 fs. The soliton states are sustained for several hours or even longer without external interference.

To explore the soliton existence range of the DKS state, we characterized the pump transmissions under various powers. When the power is at a relatively low level, the transmission is the result of simply combining the two modes (Fig. 3a). The shape of resonances changes from Lorentz to triangular when increasing the power, accompanied by the red-shift and overlap of the resonances. At high pump power, the extinction ratio of the $TE_{10}$ mode is much higher due to the low $Q$ factor, requiring more power absorption compared with the $TE_{00}$ mode. A striking soliton step (80 pm, ~10 GHz) is observed when the pump power increases to 340 mW (Fig. 3b), which is the widest soliton existence window as far as we know. When further increasing the power, the soliton step declines near linearly (~ -1.1 pm/mW) and disappears at 420 mW because of the residual thermal effects at higher power. It should be noted that the soliton existence range can be further increased to ~90 pm, when we use an adiabatic laser tuning speed (several GHz /s, tuning by hand). The slow laser tuning operation benefits from the high thermal conductance of $1.2\times10^{-2}$ W/K in our AlN microresonator (see Supplementary Fig. S1), indicating that the cavity temperature increase is 40 times less than that in a $Si_3N_4$ microresonator with similar pump power[27].

**Directly accessing the soliton state**

One highlight of our design and system is the direct access to the DKS state by simply red tuning the pump laser with one step, thereby eliminating the requirement for complex tuning or power-kick techniques. The laser was switched 50 times between two wavelengths (Fig. 4a) from off- to on-resonance states, respectively. Soliton comb operation was reliably achieved, as confirmed by monitoring the soliton spectrum and pump transmission with both OSA and real-time oscilloscope (Figs. 4c-4e), where the failed attempts are marked by red circles. The traces show that our soliton access possibility can reach as high as 100% under specific wavelength combinations. It should be noted that the step mode of tuning the pump wavelength will reduce the thermal effects within the microresonators, thus decreasing the detuning required for soliton generation compared with the results in continuous pump tuning. Moreover, the window to access the soliton comb directly is near 10 GHz ($\lambda_{start} \in$ 1550.57-1550.66 nm, $\lambda_{stop} \in$ 1550.70-1550.73 nm, see Fig. 4b), which is wide enough to make pumping with laser diode chips possible.

**Simulation results**

To better understand the soliton dynamics in our AlN microresonator, numerical simulations were carried out based on normalized Lugiato-Lefever equation (LLE)[42,43]

$$\frac{\partial E(z,\tau)}{\partial z} = -\frac{\alpha}{2}E + i\sum_{k\geq 1}\frac{\beta_k}{k!}\left(i\frac{\partial}{\partial \tau}\right)^k E + i\gamma|E|^2 E \qquad (2)$$

Where $E(z, \tau)$ is the complex electric field travelling in the cavity, $\tau$ is the time, $\alpha$ is the loss per unit length, $\gamma$ is the effective nonlinear coefficient, $\beta_k$ is the $k_{th}$-order Taylor expansion coefficient of the dispersion. Figure 5 shows the simulation results based on the LLE model for the microring studied experimentally in Fig. 2, using the simulated dispersion and estimated $Q_{int}$ (see Fig. 1e and methods). $\gamma$ is taken to be 0.82 and 0.71 $W^{-1}m^{-1}$ for $TE_{00}$ and $TE_{10}$ mode. Considering a small proportion of the pump power 140 mW for $TE_{10}$ mode, as well as the low $Q$ factor, we can only obtain a few mixing frequencies (Fig.

5a) and a Lorentzian function comb power of the effective cavity detuning (inset in Fig. 5a). For the TE$_{00}$ mode, the comb can end up in N = 4 soliton state at 210 mW. As the power increases to 400 and 600 mW, we can observe different soliton steps through the comb power curve, while the comb can end up in single soliton state, theoretically indicating that higher power is helpful for the single soliton formation. The octave-spanning single soliton spectrum has a DW around 266 THz, as shown in Fig. 5d, which agrees with the measured value 264 THz in Fig. 2d well. These results reveal that single soliton can only be obtained by pumping the TE$_{00}$ mode at high powers, while the TE$_{10}$ mode generates a narrower Kerr comb due to its relatively low $Q$ factor. By comparing the simulated and measured results, one can find that the power requirement for single soliton generation is decreased when there is the auxiliary mode near the pump mode. A more complete model including thermal effects will be developed in the future to fully explain the soliton formation processes in this novel AlN microresonator.

**Discussion**

In conclusion, we present a simple route to stably achieve the DKS in an AlN microresonator, in which a nearby auxiliary mode is slightly red-detuned from the pump mode. The auxiliary resonance can compensate for the intracavity power change and balance the thermal effects in the resonator, thus producing and broadening the soliton step significantly. A comparison of different nonlinear material platforms for single soliton generation is shown in Table 1. In this work, octave-spanning soliton microcomb ranging from 130 to 270 THz, with repetition rate of ~374 GHz, is firstly demonstrated in AlN by scanning the laser wavelength with 1 nm/s speed. The spectral bandwidth is at a similar level to that realized with the state-of-the-art Si$_3$N$_4$ technology, while more comb lines are expected in our AlN microresonators due to the smaller FSR. Moreover, the demonstrated 80-pm-wide (10 GHz) single soliton step is, to the best of our knowledge, the widest soliton accessing window so far, which will relax the strict requirements in controlling the pump or the cavity temperature. Benefiting from this, we can directly reach the soliton states with high probability by periodically switching the laser between the off- and on-resonance states with a single step. The calculated and measured pulse widths for the soliton microcomb scan are found to be as narrow as 21 and 30 fs.

In the future cavity designs, a patterned metal contact can be deposited near the microresonators for modifying the mode separation and the thermal effects in the cavity. This provides another dimension, together with optimized pump tuning speed and power, to control the thermal effect for accessing soliton states more easily. The influences of $Q$ factor and the mode separation on the soliton performance need further study. We can estimate that the pump power required for soliton generation can be decreased to tens of mW assuming the intrinsic $Q$ factor was increased to 10 million, which is needed to directly generate the soliton by pumping the passive microresonator with a laser diode chip. In addition, instead of the mixed polarizations of pump and auxiliary modes, the same polarization scheme eliminates the need of an in-line polarization controller for realizing the vision of hybrid integration. By designing different radii and waveguide widths, one can also expect the soliton microcombs with different repetition rates in different materials by pumping at other wavelength windows such as 1300 and 1064 nm. This approach provides a deterministic and simplified route to soliton modelocking in integrated optical microresonators, which is critical for the miniaturized integration and potential applications outside the laboratory.

## Methods

### Fabrication and characterization of AlN microresonator chip

The single-crystal AlN film is grown on a sapphire substrate (0001) by metal organic chemical vapor deposition (MOCVD). A 1.2-μm-thick AlN film is selected to reduce the scattering loss caused by lattice mismatch at the interface between the AlN and sapphire, while ensuring the control of geometric dispersion. An 800-nm-thick $SiO_2$ and a 70-nm-thick chromium (Cr) are deposited before spinning 900-nm-thick SPR955-0.9 photoresist. The ring resonators and bus waveguides are defined using a stepper reticle and the Nikon NSR-2005i9C Stepper System based on i-line UV illumination (365 nm Hg spectral peak). Then, a post-exposure bake is used to cure the surface roughness of the photoresist pattern. Next, the pattern is transferred to the Cr mask and the $SiO_2$ hard mask sequentially by inductively coupled plasma-reactive ion etching (ICP-RIE) using $Cl_2/O_2$ and $SF_6/Ar$, respectively. After removing the Cr mask, the $SiO_2$ mask is used to etch the AlN layer completely with an optimized $Cl_2/BCl_3/Ar$-based ICP-RIE dry etching process with the speed of ~200 nm/min. Finally, devices are encapsulated in 1.5-μm-thick PECVD-$SiO_2$ and cleaved prior to the measurement. This process provides a new route for AlN-on-sapphire fabrication, requiring simple photolithography instead of EBL as in previous reports.

The threshold power for parametric oscillation can be calculated as[34]:

$$P_{th} \approx 1.54 \frac{\pi}{2} (\frac{Q_c}{Q_L}) \frac{n^2 V_{eff}}{n_2 \lambda Q_L^2} \tag{3}$$

where $n$ is the refractive index, $V_{eff}$ is the effective modal volume, $n_2$ is the nonlinear refractive index, and $Q_c$ and $Q_L$ are the coupling and loaded quality factors of the resonators. For the AlN microresonator with FSR of ~374 GHz, around 1550 nm, we have $n \approx 2.1$, $V_{eff} \approx 644$ μm³/720 μm³ ($TE_{00}$/$TE_{10}$), $n_2 = 3.5\times10^{-19}$ m²·W⁻¹, $Q_c = 6.8\times10^5/4.5\times10^5$ ($TE_{00}$/$TE_{10}$), $Q_L = 4.8\times10^5/2.8\times10^5$ ($TE_{00}$/$TE_{10}$). Therefore, the threshold powers of $TE_{00}$ and $TE_{10}$ mode are estimated to be 39 and 145 mW, respectively.

### Setup for soliton characterization

For measuring the frequency comb, we used two lensed fibers with spot size of 2.5 μm to couple the light into the bus waveguide and collect the output. The coupling loss per facet is around ~3.2 dB. The light source used in the experiment is a tunable laser ranging from 1480 to 1640 nm (Santec TSL-710). By recording the output with a power monitor (Santec MPM210) synched with the tunable laser, we can obtain the transmission with resolution of 0.1 pm. An erbium-doped fiber amplifier (EDFA, IPG photonics, maximum output power 5W) is used to amplify the laser power for pumping the resonator. For the soliton generation in our experiment, the laser is automatically swept over the resonance (or switched between off- and on-resonance) by using laser's build-in scanning function. The output fiber including transmitted pump light and generated comb is divided into two branches, one of which is connected to the optical spectrum analyzers (YOKOGAWA AQ6375B 1200-2400 nm and AQ6370 600-1700 nm) for recording the spectrum. The other branch is injected into a fiber Bragg grating (FBG) filter to differentiate the pump transmission (band pass, BP) and generated comb sidebands (band rejection, BR) separately. The BR part is used for comb coherence characterization with electrical spectrum analyzer (ESA) and autocorrelator (APE pulseCheck). The BP part is connected to the oscilloscope (Tektronix MDO3102) and power monitor to measure the transmission traces.

## Data availability

The data sets generated and/or analysed during the current study are available from the corresponding author on reasonable request.

**Reference**

1. Diddams, S. A., Vahala, K. & Udem, T. Optical frequency combs: Coherently uniting the electromagnetic spectrum. *Science* **369**, eaay3676 (2020).

2. Diddams, S. A. et al. An optical clock based on a single trapped 199$^{Hg+}$ ion. *Science* **293**, 825 (2001).

3. Udem, T., Holzwarth, R. & Hänsch, T. W. Optical frequency metrology. *Nature* **416**, 233-237 (2002).

4. Jiang, Z., Huang, C.-B., Leaird, D. E. & Weiner, A. M. Optical arbitrary waveform processing of more than 100 spectral comb lines. *Nat. Photonics* **1**, 463-467 (2007).

5. Li, C.-H. et al. A laser frequency comb that enables radial velocity measurements with a precision of 1 cm s$^{-1}$. *Nature* **452**, 610-612 (2008).

6. Del'Haye, P. et al. Optical frequency comb generation from a monolithic microresonator. *Nature* **450**, 1214-1217 (2007).

7. T. J. Kippenberg, R. H., S. A. Diddams. Microresonator-Based Optical Frequency Combs. *Science* **332**, 6 (2011).

8. Levy, J. S. et al. CMOS-compatible multiple-wavelength oscillator for on-chip optical interconnects. *Nat. Photonics* **4**, 37 (2010).

9. Griffith, A. G. et al. Silicon-chip mid-infrared frequency comb generation. *Nat. Commun.* **6** (2015).

10. Lee, S. H. et al. Towards visible soliton microcomb generation. *Nat. Commun.* **8**, 1295 (2017).

11. Karpov, M., Pfeiffer, M. H. P., Liu, J., Lukashchuk, A. & Kippenberg, T. J. Photonic chip-based soliton frequency combs covering the biological imaging window. *Nat. Commun.* **9**, 1146 (2018).

12. Herr, T. et al. Temporal solitons in optical microresonators. *Nat. Photonics* **8**, 145 (2013).

13. Brasch, V. et al. Photonic chip–based optical frequency comb using soliton Cherenkov radiation. *Science* **351**, 357 (2016).

14. Kippenberg, T. J., Gaeta, A. L., Lipson, M. & Gorodetsky, M. L. Dissipative Kerr solitons in optical microresonators. *Science* **361**, eaan8083 (2018).

15. Suh, M.-G., Yang, Q.-F., Yang, K. Y., Yi, X. & Vahala, K. J. Microresonator soliton dual-comb spectroscopy. *Science* **354**, 600 (2016).

16. Marin-Palomo, P. et al. Microresonator-based solitons for massively parallel coherent optical communications. *Nature* **546**, 274-279 (2017).

17. Spencer, D. T. et al. An optical-frequency synthesizer using integrated photonics. *Nature* **557**, 81-85 (2018).

18. Trocha, P. et al. Ultrafast optical ranging using microresonator soliton frequency combs. *Science* **359**, 887 (2018).

19. Lucas, E. et al. Ultralow-noise photonic microwave synthesis using a soliton microcomb-based transfer oscillator.

*Nat. Commun.* **11**, 374 (2020).

20  Yi, X., Yang, Q.-F., Yang, K. Y., Suh, M.-G. & Vahala, K. Soliton frequency comb at microwave rates in a high-Q silica microresonator. *Optica* **2**, 1078-1085 (2015).

21  Guo, H. et al. Universal dynamics and deterministic switching of dissipative Kerr solitons in optical microresonators. *Nat. Phys.* **13**, 94-102 (2017).

22  Joshi, C. et al. Thermally controlled comb generation and soliton modelocking in microresonators. *Opt. Lett.* **41**, 2565-2568 (2016).

23  Bao, C. et al. Direct soliton generation in microresonators. *Opt. Lett.* **42**, 2519-2522 (2017).

24  Gong, Z. et al. High-fidelity cavity soliton generation in crystalline AlN micro-ring resonators. *Opt. Lett.* **43**, 4366-4369 (2018).

25  He, Y. et al. Self-starting bi-chromatic $LiNbO_3$ soliton microcomb. *Optica* **6**, 1138-1144 (2019).

26  Gong, Z., Liu, X., Xu, Y. & Tang, H. X. Near-octave lithium niobate soliton microcomb. *Optica* **7**, 1275-1278 (2020).

27  Li, Q. et al. Stably accessing octave-spanning microresonator frequency combs in the soliton regime. *Optica* **4**, 193-203 (2017).

28  Pfeiffer, M. H. P. et al. Octave-spanning dissipative Kerr soliton frequency combs in $Si_3N_4$ microresonators. *Optica* **4**, 684-691 (2017).

29  Lu, Z. et al. Deterministic generation and switching of dissipative Kerr soliton in a thermally controlled micro-resonator. *AIP Advances* **9**, 025314 (2019).

30  Zhou, H. et al. Soliton bursts and deterministic dissipative Kerr soliton generation in auxiliary-assisted microcavities. *Light: Sci. & Appl.* **8**, 50 (2019).

31  Zheng, Y. et al. Soliton comb generation in air-clad AlN microresonators. In *Conference on Lasers and Electro-Optics,* OSA Technical Digest (online)*,* SW4J.3 (Optical Society of America, 2020).

32  Zhang, S. et al. Sub-milliwatt-level microresonator solitons with extended access range using an auxiliary laser. *Optica* **6**, 206-212 (2019).

33  Jung, H., Xiong, C., Fong, K. Y., Zhang, X. & Tang, H. X. Optical frequency comb generation from aluminum nitride microring resonator. *Opt. Lett.* **38**, 2810-2813 (2013).

34  Liu, X. et al. Integrated high-Q xrystalline AlN microresonators for broadband Kerr and Raman frequency combs. *ACS Photonics* **5**, 1943-1950 (2018).

35  Jung, H., Stoll, R., Guo, X., Fischer, D. & Tang, H. X. Green, red, and IR frequency comb line generation from single IR pump in AlN microring resonator. *Optica* **1**, 396-399 (2014).

36  Liu, X. et al. Generation of multiple near-visible comb lines in an AlN microring via $\chi^{(2)}$ and $\chi^{(3)}$ optical nonlinearities. *Appl. Phys. Lett.* **113**, 171106 (2018).

37  Bruch, A. W. et al. Pockels soliton microcomb. *Nat. Photonics* (2020).

38  Zhang, Y. et al. Fast growth of high quality AlN films on sapphire using a dislocation filtering layer for ultraviolet light-emitting diodes. *CrystEngComm* **21**, 4072-4078 (2019).

39  Liu, J. et al. Photolithography allows high-Q AlN microresonators for near octave-spanning frequency comb and harmonic generation. *Opt. Express* **28**, 19270-19280 (2020).

40  Yi, X., Yang, Q.-F., Yang, K. Y. & Vahala, K. Theory and measurement of the soliton self-frequency shift and efficiency in optical microcavities. *Opt. Lett.* **41**, 3419-3422 (2016).

41  Karpov, M. et al. Raman self-frequency shift of dissipative Kerr solitons in an optical microresonator. *Phys. Rev. Lett.* **116**, 103902 (2016).

42  Chembo, Y. K. & Menyuk, C. R. Spatiotemporal Lugiato-Lefever formalism for Kerr-comb generation in whispering-gallery-mode resonators. *Phys. Rev. A* **87**, 053852 (2013).

43  Moille, G., Li, Q., Lu, X. & Srinivasan, K. pyLLE: A fast and user friendly Lugiato-Lefever equation solver. *J Res. Natl. Inst. Stand. Technol.* **124** (2019).

44  Liu, J. et al. Ultralow-power chip-based soliton microcombs for photonic integration. *Optica* **5**, 1347-1353 (2018).

45  Shen, B. et al. Integrated turnkey soliton microcombs. *Nature* **582**, 365-369 (2020).

## Acknowledgments

The work is supported by Science Foundation Ireland under grant number 17/NSFC/4918 and National Natural Science Foundation of China (NSFC) 61861136001.

## Author contributions

H. Z. Weng, A. Afridi, J. Liu, and J. Li carried out the device characterization. J. Liu, J. N. Dai, X. Ma, Y. Zhang, and Q. Y. Lu fabricated the devices. H. Z. Weng, J. Liu, and A. Afridi performed numerical simulations. H. Z. Weng wrote the manuscript with contribution from all authors. J. Donegan, and W. H. Guo supervised the project.

## Competing interests

The authors declare no competing financial interests.

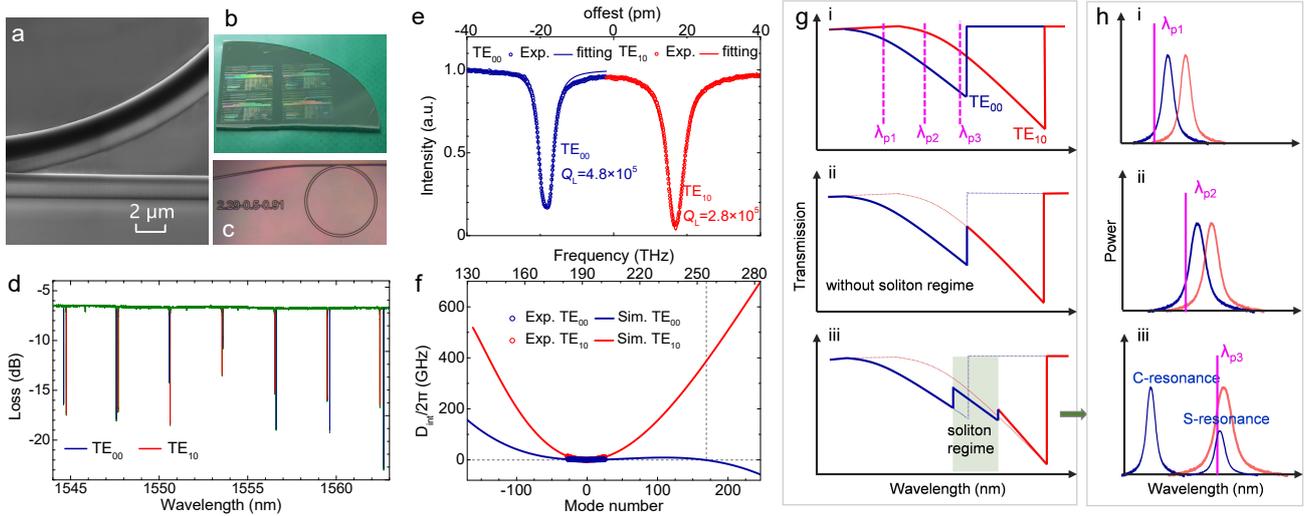

**Fig. 1 The designed AlN microring resonator with two close resonant modes near 1550 nm for accessing soliton behaviour. a** A scanning electron micrograph (SEM) image of the AlN microresonator after dry etching. **b** A photograph of a quarter of wafer fabricated with standard photolithography. **c** A microscope image of the microresonator used in this paper. **d** Transmission spectrum of the resonator, featuring two sets of resonances: the $TE_{00}$ and $TE_{10}$ modes are marked by blue and red lines. **e** Zoomed-in region of the two close resonances, spaced by ~35 pm, near 1550.7 nm with fits to determine the $Q$ values. **f** Simulated integrated dispersion $D_{int}$ in a 60-µm-radius AlN microring resonator. The cladding $SiO_2$ layer thickness is 1.5 µm. The substrate is sapphire. The circles correspond to the experimental results of around 50 resonances. Vertical dash line indicates the simulated $D_{int}$ is equal to zero at ~255 THz for $TE_{00}$ mode. **g** Schematic diagram of the pump transmission without/with soliton regime: **i**, the transmission of $TE_{00}$ and $TE_{10}$ modes are plotted separately, where three pump positions of $TE_{00}$ resonances are marked; **ii**, direct combination of the two resonance spectra without achieving soliton regime; **iii**, soliton regime realization with appropriate pump power. **h** Principle of the passive compensation of the circulating power in the microresonator by the $TE_{10}$ resonance in order to achieve soliton state shown in **iii** of (**g**): **i**, pumped at $\lambda_{p1}$ where the laser is only coupled into $TE_{00}$ mode; **ii**, pumped at $\lambda_{p2}$ where the laser is red detuned a little, where more power coupled into the $TE_{00}$ mode. $TE_{10}$ mode, which possesses only a small part of the power coupled into the cavity, will red shift due to the thermal effect. **iii**, pumped at $\lambda_{p3}$ where the laser is coupled into both modes, shows the transition into a soliton state because of the power distribution between the modes. The reduction of the pump power coupled into $TE_{00}$ mode will move the pump to the red-detuned side of the cavity resonance. In this state, the pump resonance splits into C-resonance (cavity resonance) and S-resonance (soliton resonance)[21].

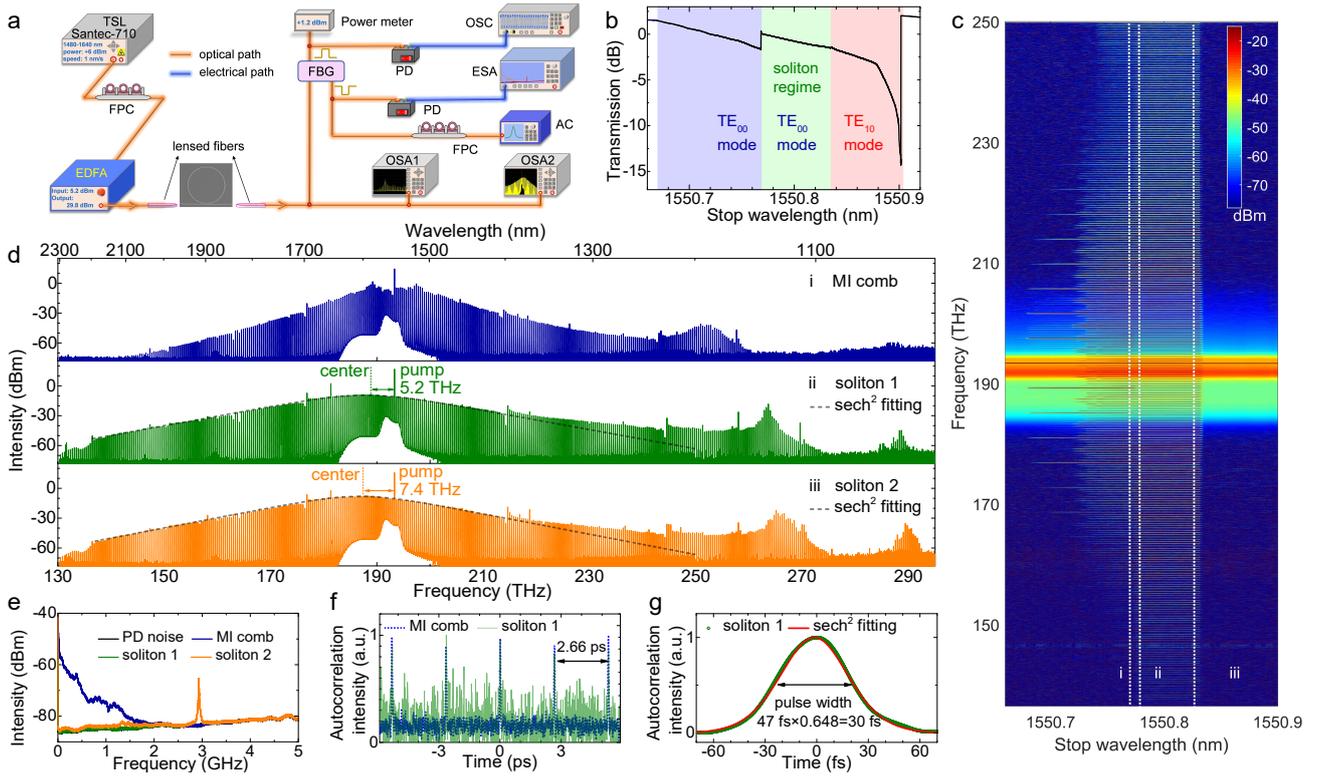

**Fig. 2 Octave-spanning dissipative Kerr soliton via pumping the two close cavity modes at 350 mW on-chip power.**
**a** Setup schematic used for soliton generation and characterization. FPC: fibre polarization controller, EDFA: erbium-doped fibre amplifier, OSA: optical spectrum analyser, AC: autocorrelator, PD: photodiode, OSC: oscilloscope, ESA: electrical spectrum analyser, FBG: fiber Bragg grating. **b** Transmission measured as the laser frequency is scanned across the resonances at 1nm/s. A soliton step with ~67 pm range for the $TE_{00}$ mode is observed, followed by the $TE_{10}$ mode. **c** Frequency comb evolution map when pump laser is scanned from the blue- to red-detuned regimes of the cavity modes at 1nm/s speed. The pump scan is conducted by maintaining the start wavelength at 1550.55 nm and increasing the stop wavelength from 1550.66 to 1550.9 nm with the step size of several pm. **d** Optical spectral snapshots of the MI comb (i) and DKS states (ii)-(iii), marked by the dash lines in (**c**). **e** Intensity noise of MI comb (blue), soliton 1 (green), soliton 2 (orange) and the noise floor of the PD used for measurements (black). **f** SHG-based autocorrelation measurement for the MI comb and soliton 1 states. **g** The autocorrelation trace for a single soliton pulse with $sech^2$ fitting, where the AC FWHM needs to be multiplied by 0.648 to yield the real pulse width.

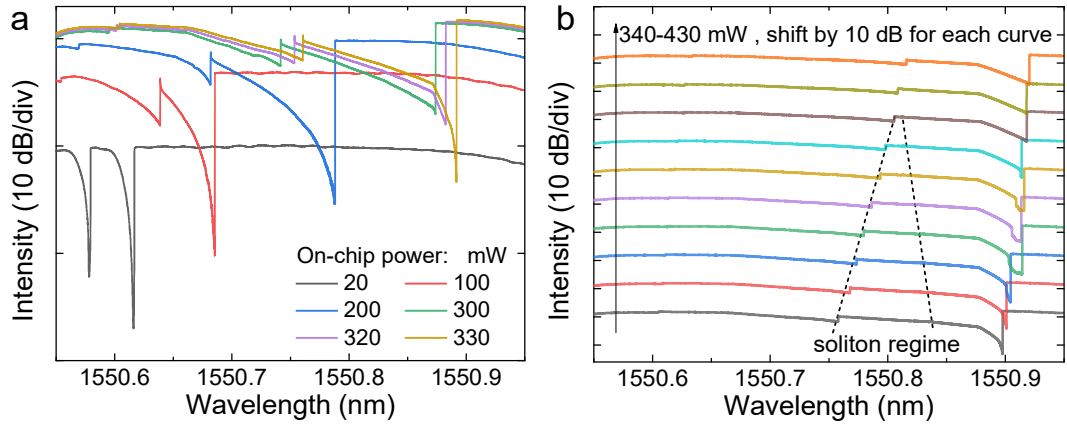

**Fig. 3 Measured pump transmissions under varied pump powers at 1nm/s laser tuning speed. a** Transmission without soliton step at 20, 100, 200, 300, 320, 330 mW on-chip power. **b** Transmission measured from 340 to 430 mW (10 mW step) on-chip power. The soliton steps observed between 340 and 410 mW are marked by the dashed lines.

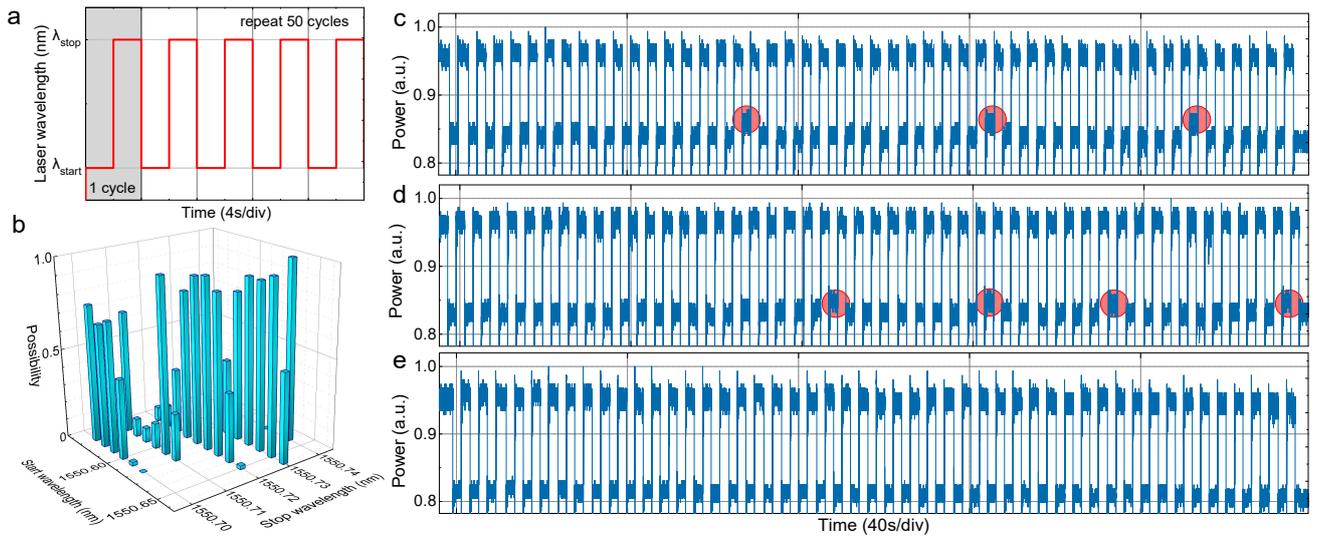

**Fig. 4 Accessing soliton states using step mode of laser wavelength tuning. a** Controlling schematic of the laser wavelength, which is switched periodically between off- ($\lambda_{start}$, 2 seconds duration) and on-resonance ($\lambda_{stop}$, 2 seconds duration) states. The switching is repeated 50 times and the spectrum is monitored by the rapid scanning in the OSA. **b** Three-dimensional histogram of the soliton accessing possibility versus the $\lambda_{start}$ and $\lambda_{stop}$. **c** Pump transmission traces recorded by oscilloscope when the $\lambda_{start}$ and $\lambda_{stop}$ are 1550.62 and 1550.72 nm, respectively. Three failed attempts (inaccessible to the soliton states) marked by red circles indicate a 94% success rate. Transmission traces when the $\lambda_{start}$ and $\lambda_{stop}$ are 1550.63 and 1550.73 nm **(d)**, and 1550.64 and 1550.74 nm **(e)**, show a 92% and 100% success rate, respectively.

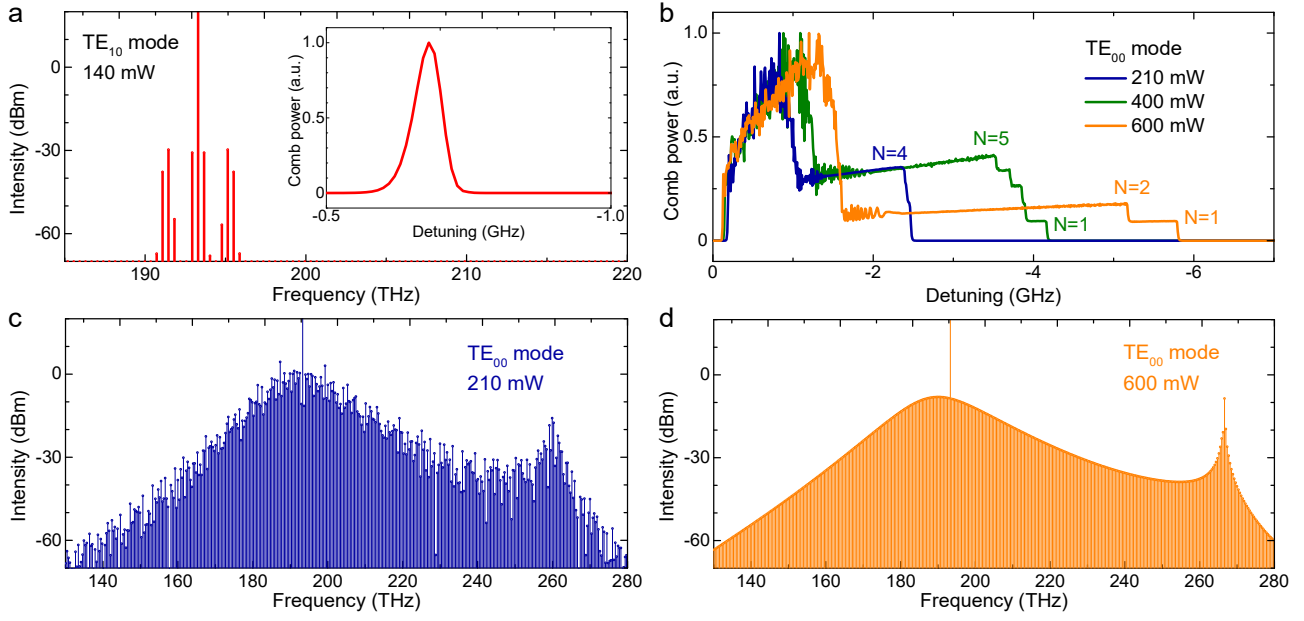

**Fig. 5 LLE simulation results. a** Simulated comb spectrum and comb power (inset) of the $TE_{10}$ mode at 140 mW pump power. **b** Simulated comb powers of $TE_{00}$ mode at the pump power of 210, 400, and 600 mW. **c** Simulated multi-soliton spectrum at 210 mW. **d** Simulated single DKS spectrum at 600 mW.

| Table 1 Performances of chip-integrated microresonators with various materials for single soliton generation | | | | | | | |
|---|---|---|---|---|---|---|---|
| Material | Cross section height×width (μm²) | FSR (GHz) | Pump mode / $D_2/2\pi$(MHz) | $Q_{load}/Q_{int}$ | Pump power (mW) | Spectral range (nm) | Tuning method; soliton step |
| $Si_3N_4$[22] | 0.95×1.5 | 200 | -/- | -/- | 71 | 1470-1620 | Thermal controlling via an integrated heater |
| $Si_3N_4$[23] | 0.8 × 2 | ~230 | -/- | $1.4×10^6$/- | 200 | 1400-1700 | Forward[b] and backward[c] tuning |
| $Si_3N_4$[27] | 0.62×1.75 | 1000 | $TE_{00}$/53±2 | $4×10^5/2×10^6$ | 120±15 | 1100-2320[a] | Forward tuning -100 GHz/s; ~2 pm |
| $Si_3N_4$[28] | 0.74×1.425 | 1000 | $TE_{00}$/49.7 | $1×10^6$/- | 455 | 1100-2300[a] | Forward and backward tuning |
| $Si_3N_4$[44] | 0.81×1.58 | 99 | $TE_{00}$/1.23 | -/$15×10^6$ | 6.2 | 1540-1620 | Piezo laser tuning |
| $Si_3N_4$[45] | - | 40 | -/- | -/$16×10^6$ | 30 | 1520-1600 | Integrated with DFB laser, turn-key soliton |
| $LiNbO_3$[25] | 0.41×2 | 199.7 | $TE_{00}$/1.76 | $2.2×10^6$/- | 33 | 1470-1650 | Forward or backward tuning |
| $LiNbO_3$[26] | 0.56×1.45 | 335 | $TE_{00}$/0.41 | ~$5×10^5/1×10^6$ | 240 | 1190-2140 | Backward tuning 62.5 GHz/s; ~0.5 GHz |
| AlN[24] | 1×1.3 | ~560 | $TE_{00}$/6.37 | $6.5×10^5$ /$9.3×10^5$ | 620 | 1400-1700 | Single-sideband modulation, forward tuning 60 pm/μs |
| AlN[31] | 0.8×3 | 225 | $TE_{00}$/- | -/$2.4×10^6$ | ~1000 | 1400-1700 | Dual-pump, forward tuning 20 nm/s / 0.2 pm |
| AlN this work | 1.2×2.29 | 374 | $TE_{00}$/4.8 | $4.8×10^5$ /$1.6×10^6$ | 350 | 1100-2300[a] | With auxiliary mode, one-step or forwarding tuning 1nm/s; **~80 pm (10 GHz)** |

[a] Represents the octave-spanning spectral range; [b] Represents increasing of the pump wavelength; [c] Represents decreasing of the pump wavelength; DFB, distributed feedback.

# Supplementary Information - Directly accessing octave-spanning dissipative Kerr soliton frequency combs in an AlN microring resonator

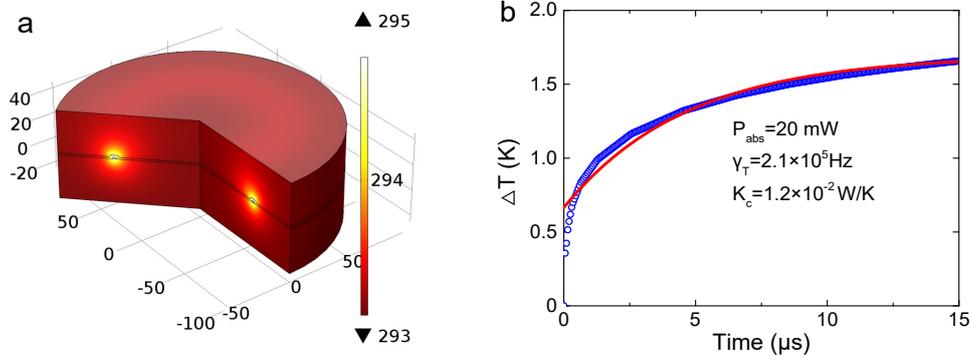

Supplementary Figure 1. **Numerical simulation of the thermal properties in a 60-μm-radius AlN microring resonator with waveguide width of 2.29 μm, height of 1.2 μm, and sidewall angle of 72°. a** Steady-state temperature distribution with 20 mW absorption power. **b** Simulated resonator temperature transformation (blue circles) and the exponential fitting (red line).

The thermal dynamics in a microresonator are described by the following equation [1-3]:

$$\frac{dT_{eff}}{dt} = -\gamma_T (T_{eff} - \frac{P_{abs}}{K_c}) \quad \text{(S1)}$$

where $T_{eff}$ is the effective cavity temperature, $\gamma_T$ is the thermal decay rate, $K_c$ is the thermal conductance of the microresonator at the absorbed optical power $P_{abs}$. By solving the heat transfer equation based on the finite element method (FEM), we analysis the thermal dynamics and temperature distributation (Fig. S1a) for a 60-μm-radius AlN microring resonator. By fitting the simulation results (Fig. S1b) with Eq. S1, we can extract the $\gamma_T = 2.1 \times 10^5$ Hz and $K_c = 1.2 \times 10^{-2}$ W/K. The thermal effects in this AlN microresonator occur much more quickly than in $Si_3N_4$ due to the sapphire substrate in the AlN case.

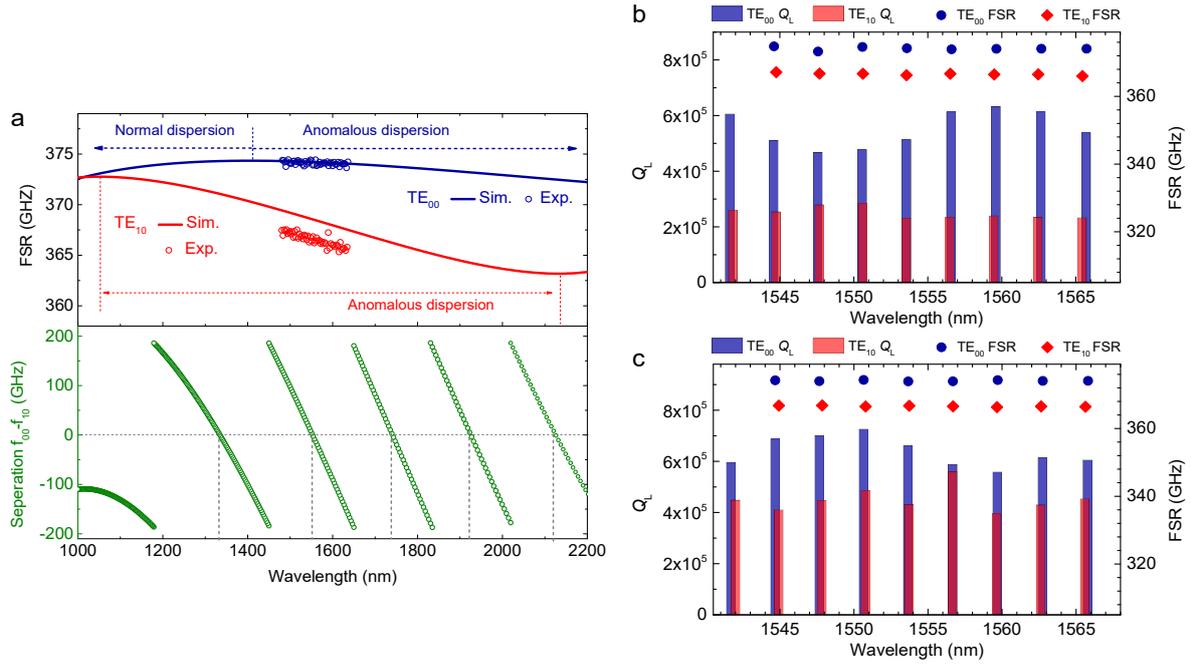

Supplementary Figure 2. **Resonator design for realization of two close TE modes. a** FEM-based simulation of FSR (upper) and the separation (lower) between $TE_{00}$ modes and the nearest $TE_{10}$ modes in and AlN microring resonator with the same parameters used in Fig. **S1a**. The normal and anomalous dispersion regimes for the two modes are marked by the dash lines and arrows. Experimental load $Q$ factors and FSRs for **b** device 1 (used in main text) and **c** device 2 (used for Fig. **S4**) with the same geometry.

In order to realize two nearby TE modes around 1550 nm in the experiment, we firstly simulated the eigenfrequencies for the fundamental transverse-electrical ($TE_{00}$) modes and first-order TE ($TE_{10}$) modes by FEM. The free spectral range (FSR, Fig. S2a upper) of $TE_{00}$ modes increases from 372.49 to 374.35 GHz with the wavelength increasing from 1000 to 1406 nm, and then decreases to 372.24 GHz at ~2196 nm, which correspond to the normal and anomalous dispersion, respectively. The FSR of $TE_{10}$ modes decreases from 372.77 to 363.18 GHz with the wavelength increasing from 1052 to 2137 nm (anomalous dispersion), while the large decrease in amplitude corresponds to the large curvature of integrated dispersion (see Fig. 1f). We can find that $TE_{00}$ modes have large FSRs in the calculated range and smaller changing amplitude, which is helpful for the modelocking during soliton generation. The frequency separation between $TE_{00}$ modes and the nearest $TE_{10}$ modes are also calculated and extracted. The two mode families can be close to each other at 1331.77, 1550.98, 1739.48, 1922.84, and 2120.63 nm (Fig. S2a lower). Specially, for the target C-band, the operating position is ~1551 nm and the $TE_{00}$ is 3.3 GHz (26 pm) higher than the nearby $TE_{10}$ mode in frequency.

The designed microresonators were characterized through transmission measurements at a low injection power of ~40 µW. For both resonators shown in Figs. S2b-2c, the FSRs of $TE_{00}$ and $TE_{10}$ modes near 1550 nm are ~374 and ~366 GHz respectively. For device 1 (Fig. S2b, also shown in Fig. 2), two modes are close near 1550.7 nm, and the auxiliary $TE_{10}$ mode is red detuned from the pump $TE_{00}$ mode slightly (separation is 35 pm, ~4.4 GHz), which agrees with the simulation results very well. The average loaded $Q$ factors for the $TE_{00}$ and $TE_{10}$ modes are ~$5.5 \times 10^5$ and ~$2.3 \times 10^5$, respectively, while their minimum and maximum values of $4.8 \times 10^5$ and $2.8 \times 10^5$ are observed near the meeting. Here, the mode crossing is avoided due to the different field distributions of the two mode families. For device 2 (Fig. S2c), the two modes are close to 1556.7 nm with a separation of 12 pm (~1.5 GHz). The average loaded $Q$ factors for the $TE_{00}$ and $TE_{10}$ modes are ~$6.5 \times 10^5$ and ~$4.5 \times 10^5$, respectively. The differences between the two devices can be explained by the fabrication variation, such as the ring width or angle deviation. These experimental results imply that our design is easy to achieve and repeatable in experiment using standard photography and dry etching.

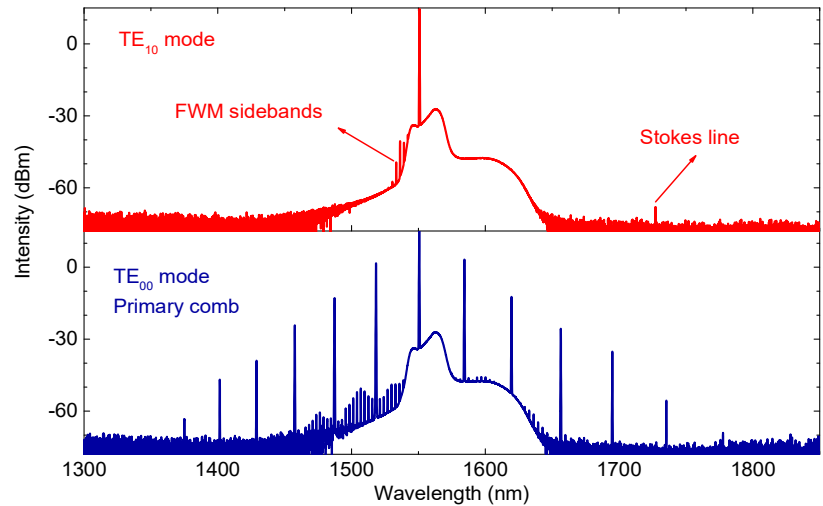

Supplementary Figure 3. **Measured comb spectra for device 1 when pump the $TE_{00}$ and $TE_{10}$ modes, at 350 mW.** The FWM sidebands and Stokes lines are indicated by the arrows when pumping the $TE_{10}$ mode.

In the transmission at high pump power shown in Fig. 2b, the $TE_{10}$ mode has higher extinction ratio, indicating that more power needs to be absorbed to attain the parameter oscillation threshold and produce new mixing frequencies. Therefore, a spectrum with a few lines caused by four-wave mixing (FWM) and stimulated Raman scattering are observed (Fig. S3 upper), which is extracted from Fig. 2c at the stop wavelength of 1550.89 nm before dropping out of the auxiliary $TE_{10}$ mode. The primary comb spectrum of $TE_{00}$ mode is also extracted and plotted, which has twelve sidebands with line spacing of 11 FSR (Fig. S3 lower).

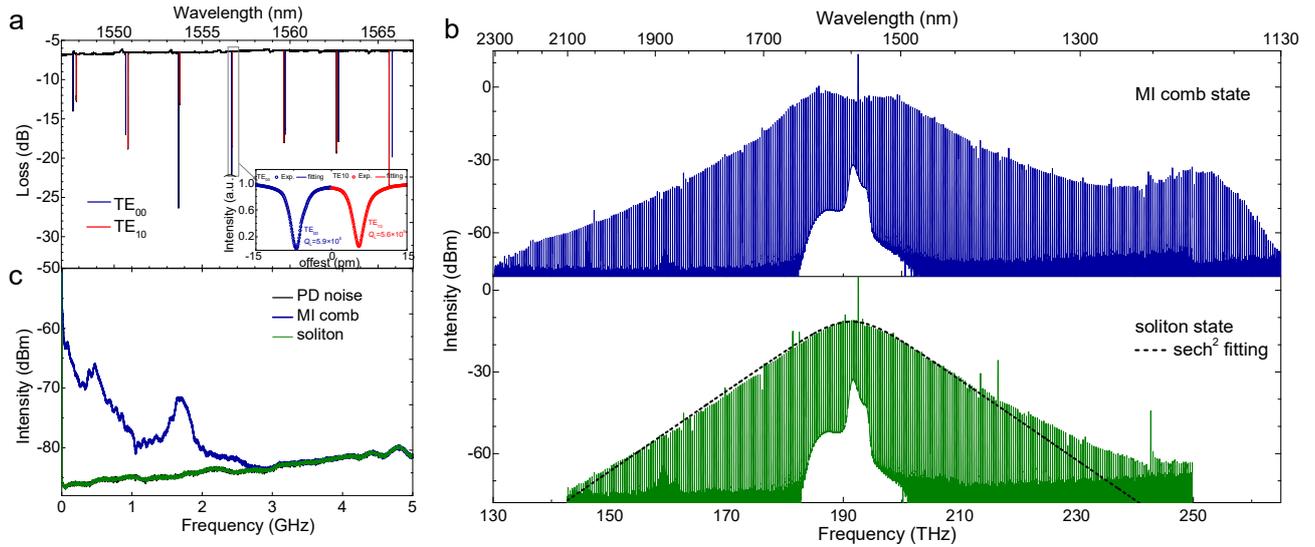

Supplementary Figure 4. **Near octave-spanning soliton demonstrated in device 2 at 400 mW on-chip power. a** Transmission spectrum of the resonator. Inset shows the Lorentz fitting of the two close resonances, spaced by ~12 pm, near 1556.7 nm. **b** Measured spectra of MI comb state (upper) and soliton state (lower, laser tuning speed 14 nm/s). **c** Intensity noise of MI comb state and soliton state and the noise floor of the PD used for measurements.

In another experiment, we demonstrate a near octave-spanning soliton with device 2, which has a similar transmission as device 1, while the $TE_{00}$ resonance is now only 12 pm shorter in wavelength than the $TE_{10}$ resonance around 1556.7 nm (Fig. S4a). The $Q_L$ for the $TE_{00}$ and $TE_{10}$ modes are $5.9 \times 10^5$ and $5.6 \times 10^5$ (inset of Fig. S4a, also see Fig. S1c), respectively. By sweeping the laser wavelength from blue- to red-detuning, we can observe the modulation instability (MI) comb and single soliton state (Fig. S4b), with manual operation and 14 nm/s sweeping speed, respectively. By fitting the soliton spectrum with a $sech^2$ shape (dash line), we can find the 3-dB bandwidth is ~9 THz, corresponding to 24 modes. The low frequency intensity-noise (Fig. S4c) confirms the coherence and modelocking of the soliton state. Here, the $TE_{10}$ mode is much closer to the $TE_{00}$ mode in this case and receives a large proportion of the pump power, which will result in the narrower soliton spectral bandwidth compared with device 1. Also, to reach the soliton state, we utilize a faster pump tuning speed to avoid the influence of the nonlinear frequency conversion from the $TE_{10}$ mode because of the narrower mode separation.

## Supplementary References

1       Il'chenko, V., & Gorodetskii, M. Thermal nonlinear effects in optical whispering gallery microresonators. *Laser Phys* **2**, 1004–1009 (1992).

2       Carmon, T., Yang, L., & Vahala, K. J. Dynamical thermal behavior and thermal self-stability of microcavities. *Opt.*

*Express* **12**, 4742–4750 (2004).

3  Li, Q. et al. Stably accessing octave-spanning microresonator frequency combs in the soliton regime. *Optica* **4**, 193-203 (2017).